\documentclass[10pt,journal,compsoc]{IEEEtran}

\usepackage[utf8]{inputenc}
\usepackage[T1]{fontenc}
\usepackage{graphicx}
\usepackage{amsmath,amssymb}
\usepackage{booktabs}
\usepackage{array}
\usepackage{multirow}
\usepackage{url}
\usepackage{xcolor}
\usepackage{tikz}
\usetikzlibrary{arrows.meta,positioning,shapes.geometric,calc,fit,backgrounds}
\usepackage{hyperref} 
\usepackage{siunitx}
\hypersetup{colorlinks=true, linkcolor=black, filecolor=black, urlcolor=cyan}
\usetikzlibrary{arrows.meta,positioning,calc,fit,backgrounds}

\newcommand{\rc}[1]{\textcolor{red!70!black}{#1}}

\begin{document}
\pagestyle{empty}
\title{Short-Form Video Exposure and Academic Strain: A Dose-Response Analysis and Acquiescence Diagnosis}

\author{%
  \IEEEauthorblockN{Md.~Alamgir~Hossain\textsuperscript{\textcolor{red}{1,\,*}}}\\[2pt]
  \IEEEauthorblockA{%
   \textsuperscript{\textcolor{red}{1}}Skill Morph Research Lab., Skill Morph, Dhaka, Bangladesh\\
    \textsuperscript{\textcolor{red}{1}}Department of Computer Science and Engineering,
    State University of Bangladesh, Dhaka, Bangladesh\\[3pt]
    \textsuperscript{\textcolor{red}{1}}E-mail:~\href{mailto:alamgir.cse14.just@gmail.com}{alamgir.cse14.just@gmail.com}, ORCID:~%
    \href{https://orcid.org/0000-0001-5120-2911}{0000-0001-5120-2911}\\[3pt]
    \textsuperscript{\textcolor{red}{*}}Corresponding Author: Md.~Alamgir~Hossain
    (E-mail:~\href{mailto:alamgir.cse14.just@gmail.com}{alamgir.cse14.just@gmail.com};~%
    ORCID:~\href{https://orcid.org/0000-0001-5120-2911}{0000-0001-5120-2911})%
  }%
}

%\markboth{Preprint, July 2026}{Hossain: Short-Form Content Exposure, Attention Fragmentation, and Acquiescence Bias}

\IEEEtitleabstractindextext{%
\begin{abstract}
Short-form video platforms now mediate a large share of undergraduate leisure time, prompting concern that compulsive, high-arousal consumption erodes the sustained attention required for deep learning while accelerating academic burnout. A recurring weakness in this fast-growing literature is that most evidence rests on single-source Likert instruments whose validity is rarely interrogated for response-style contamination. The present work analyses 863 university students who completed a 67-item, seven-construct battery covering short-form content addiction, attention fragmentation, notification overload, cognitive fatigue, deep learning engagement, academic burnout, and self-regulation, alongside categorical exposure indicators. A layered analytic protocol progressing from descriptive screening to dimensionality assessment, reliability, convergent validity, careless-responding diagnostics, and known-groups criterion validity was implemented in Python. Two findings emerge and reinforce one another. First, self-reported daily exposure and in-study checking frequency exhibit robust, monotone dose-response gradients on every maladaptive construct, with addiction showing the strongest effect ($\eta^2=0.117$, $\rho=0.273$, $p<0.001$), confirmed by nonparametric tests. Second, the intended latent measurement model collapses: parallel analysis retains a single factor per construct, average variance extracted never exceeds 0.31, and a general agreement factor explains 59\% to 89\% of construct variance, reversing the theoretically negative addiction-to-deep-learning association from $+0.55$ to $-0.49$ once partialled. The contribution is therefore twofold: a replicable demonstration that exposure predicts cognitive-academic strain, and a diagnostic account showing that acquiescence can mask and invert those very relationships. The results caution that unbalanced, uniformly worded scales jeopardise inference in behavioural-addiction research and argue for balanced instrumentation, embedded quality controls, and explicit response-style modelling.
\end{abstract}

}

\maketitle
\IEEEdisplaynontitleabstractindextext
\IEEEpeerreviewmaketitle
%=====================================================================
\section{Introduction}
%=====================================================================

\subsection{Background and Motivation}
Short-form video has become the dominant grammar of digital attention among young adults, delivered as rapid, algorithmically curated streams on TikTok, Instagram Reels, Facebook Reels, and YouTube Shorts. The format is engineered around immediate reward, continuous novelty, and frictionless progression from one clip to the next, and its saturation of student leisure time has reframed a longstanding question in higher education \cite{roberts2025technology, yin2024virality}. The concern is no longer whether undergraduates use these platforms, which is now near universal, but whether habitual and compulsive consumption reshapes the very faculties on which academic work depends: the capacity to sustain attention, to process complex material deeply, and to regulate effort under pressure. That question carries genuine stakes, because deep and effortful engagement is the operative currency of university learning, and any systematic erosion of it propagates into comprehension, retention, and eventual attainment. The motivation for the present work therefore lies not in cataloguing another correlate of screen time, but in clarifying how, and how reliably, patterns of short-form consumption map onto the cognitive and affective machinery of student learning \cite{hsieh2025smartphone, firth2024online, clemente2024digital}.

\subsection{Problem Statement}
The empirical case connecting short-form consumption to academic cognition is expanding rapidly, yet it rests on a fragile methodological foundation. A large share of recent studies administers a single self-report questionnaire at one time point, correlates its subscales, and interprets the resulting coefficients as evidence of substantive psychological relationships \cite{li2024shortform,haliti2024reels,shuja2025mindfulness}. Designs of this kind are exposed to common method variance and, more insidiously, to acquiescence, the disposition to endorse items irrespective of their content \cite{baumgartner2001response,primi2020acquiescence}. When an entire battery is worded in a single evaluative direction, this stylistic tendency does not merely add noise; it inflates internal consistency, distorts factor structure, and can amplify or even reverse the associations that anchor a study's conclusions \cite{maydeu2006randomintercept,martins2024comparison}. The problem, stated plainly, is that a literature drawing increasingly strong inferences about the cognitive costs of short-form media has largely proceeded as if these measurement threats were immaterial, leaving open the possibility that some of its headline findings reflect how respondents answer rather than what they experience.

\subsection{Research Gap}
Two distinct deficiencies converge to define the gap this study addresses. Substantively, few investigations triangulate self-reported psychological constructs against behavioural exposure indicators that are comparatively resistant to a uniform agreement style, so it remains unresolved how much of the reported association reflects a genuine dose-response structure as opposed to shared method variance. Methodologically, the response-style vulnerabilities of the newly authored short-form-addiction batteries now circulating in this domain have not been examined systematically, even though balanced instrumentation and explicit acquiescence modelling are well established remedies elsewhere in psychometrics \cite{maydeu2006randomintercept,savalei2014robustness}. The two literatures have developed in isolation: applied behavioural-addiction research seldom deploys acquiescence diagnostics, while the corrective psychometric literature rarely interrogates the specific instruments proliferating in that applied space. Bridging this separation within a single, adequately powered dataset, by auditing a representative multidimensional instrument for precisely the distortions theory predicts and by testing whether behavioural exposure recovers the substantive gradient that contaminated self-report obscures, is the contribution the field currently lacks.

\subsection{Objectives}
Three objectives structure the investigation and follow directly from the gap identified above:
\begin{itemize}
\item[\textbf{O1.}] To quantify whether graded, self-reported exposure to short-form platforms predicts a coherent profile of maladaptive cognitive and academic outcomes, using criterion-oriented comparisons that are robust to a uniform agreement style.
\item[\textbf{O2.}] To subject the seven-construct self-report battery to a complete measurement-model audit, examining in sequence its distributional properties, internal consistency, dimensionality, and convergent validity rather than presuming these prerequisites hold.
\item[\textbf{O3.}] To determine, should the audit fail, whether acquiescence is the operative mechanism, and to characterise the magnitude and direction of the distortion it imposes on the target relationships.
\end{itemize}
Pursued jointly, these objectives convert a conventional survey into a controlled test of what such surveys can legitimately support, ensuring that every inference the study advances is licensed by evidence accumulated in sequence rather than presumed at the outset.

\subsection{Hypothesis Development}
The substantive predictions derive from an integrative frame that unites three complementary accounts. The Interaction of Person, Affect, Cognition, and Execution model explains why algorithmically reinforced consumption becomes compulsive and locates weakened executive control as the switch that converts urges into behaviour \cite{brand2016ipace}. The limited-capacity account of mediated processing explains why continual, high-arousal switching should degrade the sustained and effortful processing that deep learning demands \cite{lang2000limited}. Conservation of Resources theory explains why repeated depletion without recovery should culminate in the exhaustion, disengagement, and reduced efficacy that constitute academic burnout \cite{hobfoll1989cor}, while the strength model of self-control motivates self-regulation as a boundary condition on the earliest link in the chain. From this synthesis, six hypotheses were advanced:

\begin{itemize}
\item[\textbf{H1.}] Exposure to short-form content is positively associated with short-form content addiction, attention fragmentation, notification overload, cognitive fatigue, and academic burnout, such that heavier consumers report a more pronounced maladaptive profile.
\item[\textbf{H2.}] Exposure is negatively associated with the adaptive constructs, namely deep learning engagement and self-regulation.
\item[\textbf{H3.}] Short-form content addiction is positively associated with the proximal mechanisms: (\textbf{H3a}) attention fragmentation and (\textbf{H3b}) cognitive fatigue.
\item[\textbf{H4.}] Notification overload is positively associated with (\textbf{H4a}) attention fragmentation and (\textbf{H4b}) cognitive fatigue, operating as a second antecedent of attentional strain.
\item[\textbf{H5.}] The proximal mechanisms predict the academic outcomes, such that attention fragmentation and cognitive fatigue are (\textbf{H5a}) negatively associated with deep learning engagement and (\textbf{H5b}) positively associated with academic burnout, and (\textbf{H5c}) jointly transmit the influence of addiction onto those outcomes.
\item[\textbf{H6.}] Self-regulation moderates the addiction-to-fragmentation pathway, attenuating it among students with stronger regulatory capacity.
\end{itemize}
Collectively, H1 through H6 specify a coherent nomological network extending from behavioural exposure through compulsive engagement and attentional mechanisms to divergent learning and wellbeing outcomes. The empirical viability of that network, as the analysis will demonstrate, is contingent on a measurement structure capable of separating these conceptually adjacent constructs, which is precisely the premise the audit is designed to interrogate.

\subsection{Contributions}
The study makes four contributions that jointly distinguish it from the prevailing survey tradition in this area. It furnishes a dose-response characterisation of exposure and cognitive-academic strain that does not depend on the fragile inter-item correlations dominating the field, thereby offering criterion evidence that survives response-style contamination. It supplies a transparent and fully reproducible measurement audit demonstrating that a uniformly worded battery can present enviable reliability while failing convergent validity in a large sample. It isolates acquiescence as the operative cause of that failure and shows quantitatively that the response style reverses the sign of the field's central relationship, converting an apparent positive association between addiction and deep learning into the theoretically expected negative one once the agreement component is partialled. Finally, it translates these diagnostics into concrete instrumentation guidance that would immunise subsequent work against the same failure. The recurring novelty across these contributions is the reframing of a measurement artefact, ordinarily relegated to a limitations paragraph, into a substantive and generalisable finding about the reliability of an entire class of studies.

\subsection{Paper Structure}
The remainder of the paper is organised as follows. Section II reviews the applied and psychometric literatures and formalises the gap at their intersection. Section III details the research design, sample, instrument, preprocessing, and the statistical apparatus, with each procedure defined by its governing notation, and presents the a priori analytical framework. Section IV reports the layered results, progressing from sample description and distributional screening through the measurement audit and the acquiescence diagnosis to the behavioural dose-response evidence. Section V interprets these findings, explaining why the observed pattern arises, situating it against prior work, and drawing out its theoretical and practical implications. Section VI concludes by consolidating the contributions, acknowledging the limitations, and specifying an agenda for balanced, longitudinally informed replication.

%=====================================================================
\section{Literature Review}
%=====================================================================
Research on short-form video and academic functioning has accelerated markedly, and the modal design correlates an addiction scale with academic or attentional outcomes in a single cross-sectional survey. Representative recent work reports that short-form video app addiction predicts higher academic anxiety and lower academic engagement, with mindfulness partially mediating both paths in a large undergraduate sample analysed by structural equation modelling \cite{li2024shortform}. Parallel studies link problematic short-form use to shortened attention span and impaired academic performance \cite{haliti2024reels}, to cognitive failures alongside diminished mindfulness \cite{shuja2025mindfulness}, and to weakened attention control and delayed gratification in young adults \cite{singh2025attention}. Systematic syntheses of adolescent and youth samples echo a consistent, if correlational, pattern of cognitive cost \cite{review2026shortvideo}.

A comparative reading of this evidence reveals convergence at the level of direction and divergence at the level of rigour. Studies agree that heavier or more compulsive use accompanies poorer attentional and academic indicators, and several extend the picture to mediating mechanisms such as procrastination and self-efficacy \cite{shortvideo2024physical}. Yet the instruments differ, the constructs overlap conceptually, and the analytic burden is placed almost entirely on latent models estimated from Likert subscales whose discriminant validity is seldom demonstrated. When multiple constructs are measured with same-direction items and then entered into a covariance-based model, the resulting fit statistics can be satisfied by shared method variance rather than by the hypothesised theoretical structure, a possibility that is rarely tested and almost never ruled out.

The psychometric literature makes the nature of that vulnerability precise. Acquiescence is a stable respondent tendency to endorse items regardless of content, and its effect on measurement is systematic rather than random: it inflates reliability, biases factor loadings, and, critically, increases correlations among items sharing a semantic pole while decreasing correlations among items of opposite poles \cite{primi2020acquiescence,baumgartner2001response}. In an entirely positively keyed multi-scale battery, this mechanism manufactures the appearance of coherent constructs and can convert theoretically negative relationships into positive ones. The standard remedies are well established. Balanced scales with reverse-keyed items allow acquiescence to be separated from content, and model-based corrections such as the random-intercept item factor analysis add a single agreement factor whose loadings are fixed across all items \cite{maydeu2006randomintercept,savalei2014robustness}. Comparative simulation work indicates that these corrections perform well when their assumptions are met but that the recovery of acquiescence-free scores is not guaranteed under unbalanced designs \cite{martins2024comparison}.

Critically, the two literatures have not been brought into contact. Behavioural-addiction studies of short-form video rarely deploy balanced instrumentation or report acquiescence diagnostics, while the psychometric corrective literature seldom examines the specific instruments now proliferating in that applied domain. The consequence is a body of applied findings whose internal validity is largely assumed rather than demonstrated, and a body of methodological knowledge whose warnings have not been operationalised where they are most needed. The present study addresses that disconnection directly by auditing a representative multidimensional short-form battery for the precise distortions the psychometric literature predicts, and by testing whether behavioural exposure indicators, which are less exposed to a uniform agreement style, recover the substantive gradient that the contaminated self-report correlations obscure. This dual strategy defines the gap the study fills.

%=====================================================================
\section{Methodology}
%=====================================================================
\subsection{Research Design}
A cross-sectional, correlational survey design was adopted, with a deliberately layered analytic protocol that treats measurement validity as an object of investigation rather than a precondition assumed to hold. The protocol proceeds in stages, each conditioning the next: descriptive and distributional screening, internal-consistency reliability, dimensionality assessment, convergent validity, careless-responding diagnostics, response-style modelling, and finally known-groups criterion validity against behavioural exposure. Structuring the analysis this way ensures that any inference drawn at a later stage is licensed by the evidence accumulated at earlier ones, and it makes the eventual measurement failure both visible and diagnosable rather than hidden inside a single global model.

\subsection{Participants and Sampling Justification}
Responses were obtained from university students through an anonymous online questionnaire that opened with an informed-consent item and eligibility statement requiring current enrolment and a minimum age of sixteen years. Of 867 submissions, four respondents reported an age below the stated threshold and were removed to preserve a clean eligibility criterion, yielding an analytic sample of $N=863$. The sample is balanced by gender (49.9\% male, 50.1\% female) and concentrated in the 21-to-26 age band (80.2\%), with private-university students predominating (73.1\%). Undergraduates constitute the largest academic level (40.0\%) and Engineering or Technology the largest discipline (32.9\%). This composition supports adequate statistical power for the planned comparisons while implying that generalisation beyond younger, predominantly private-university cohorts should be made cautiously. Table~\ref{tab:sample} summarises the sample.

\subsection{Instrument}
The battery comprised 67 items rated on a five-point Likert scale from strongly disagree (1) to strongly agree (5), grouped a priori into seven constructs: short-form content addiction (SFCA, 16 items), attention fragmentation (AF, 12), notification overload (NO, 5), cognitive fatigue (CF, 5), deep learning engagement (DLE, 12), academic burnout (AB, 12), and self-regulation (SR, 5). Four constructs (SFCA, AF, DLE, AB) were specified as higher-order, with three or four theorised facets each. Two categorical behavioural indicators, average daily time on short-form platforms and frequency of checking such platforms during study time, served as exposure measures. Every item was worded in the same evaluative direction, a design feature whose consequences the analysis later exposes.

\subsection{Data Preprocessing}
Let $\mathbf{X}\in\mathbb{R}^{n\times p}$ denote the item response matrix with $n=863$ respondents and $p=67$ items, entry $x_{ij}\in\{1,2,3,4,5\}$. Column labels exported from the survey platform carried trailing whitespace and full question text; each item column was renamed to its canonical code (for example, ``SFCA1. I often open short-form content apps\ldots'' to \texttt{SFCA1}) by extracting the token preceding the first period. Completeness was verified as
\begin{equation}
\textstyle\sum_{i=1}^{n}\sum_{j=1}^{p}\mathbb{1}\!\left[x_{ij}=\varnothing\right]=0,
\end{equation}
and no duplicate response vectors were present. The eligibility filter removed records with reported age below sixteen, $\mathcal{D}=\{i:\text{age}_i<16\}$, $|\mathcal{D}|=4$, giving the analytic set $\{1,\ldots,n\}\setminus\mathcal{D}$.

For a construct $c$ with item index set $I_c$, the composite score for respondent $i$ was the item mean
\begin{equation}
\bar{X}_{ic}=\frac{1}{|I_c|}\sum_{j\in I_c} x_{ij},
\label{eq:composite}
\end{equation}
and a person-level acquiescence, or agreement, index was defined across all items as
\begin{equation}
a_i=\frac{1}{p}\sum_{j=1}^{p} x_{ij}.
\label{eq:ars}
\end{equation}
All items shared a common keying direction, so no reverse scoring was applied; this was confirmed empirically through corrected item-total correlations reported below.

\subsection{Statistical Approaches}
Each procedure is defined in the order applied.

\emph{Distributional screening.} For each item, skewness and excess kurtosis were computed and evaluated against the thresholds $|\text{skew}|<2$ and $|\text{kurtosis}|<7$ commonly used to flag severe non-normality \cite{curran1996normality}, motivating estimators that do not assume multivariate normality.

\emph{Internal consistency.} Reliability was quantified by Cronbach's alpha
\begin{equation}
\alpha=\frac{k}{k-1}\!\left(1-\frac{\sum_{j=1}^{k}\sigma_j^2}{\sigma_T^2}\right),
\label{eq:alpha}
\end{equation}
with $k$ items, item variances $\sigma_j^2$, and total-score variance $\sigma_T^2$ \cite{cronbach1951alpha}, and by McDonald's omega
\begin{equation}
\omega=\frac{\left(\sum_{j=1}^{k}\lambda_j\right)^2}{\left(\sum_{j=1}^{k}\lambda_j\right)^2+\sum_{j=1}^{k}\left(1-\lambda_j^2\right)},
\label{eq:omega}
\end{equation}
where $\lambda_j$ are standardised loadings from a single-factor solution \cite{mcdonald1999test}. The corrected item-total correlation guarded against miskeyed items.

\emph{Dimensionality.} Factorability was assessed by the Kaiser-Meyer-Olkin index \cite{kaiser1974kmo} and Bartlett's test of sphericity. The number of factors retained was determined by Horn's parallel analysis \cite{horn1965parallel}, which compares observed eigenvalues $\ell_m$ against the 95th percentile of eigenvalues from random data of identical dimensions, retaining factor $m$ when $\ell_m>\ell_m^{\text{PA(95)}}$. Exploratory factor analysis with oblique (promax) rotation, forced to the theorised number of factors, tested whether items recovered their assigned facets.

\emph{Convergent validity.} For each construct, average variance extracted and composite reliability were computed from standardised loadings as
\begin{equation}
\text{AVE}=\frac{1}{k}\sum_{j=1}^{k}\lambda_j^2,\qquad
\rho_c=\frac{\left(\sum_{j=1}^{k}\lambda_j\right)^2}{\left(\sum_{j=1}^{k}\lambda_j\right)^2+\sum_{j=1}^{k}\left(1-\lambda_j^2\right)},
\label{eq:ave}
\end{equation}
with the conventional thresholds $\text{AVE}\geq0.50$ and $\rho_c\geq0.70$ \cite{fornell1981ave}. A guided item-retention loop iteratively removed the weakest-loading indicator until the thresholds were met or a floor of three items was reached.

\emph{Careless responding.} Three indices screened for inattentive answering \cite{meade2012careless}: the longstring (longest run of identical consecutive responses), the intra-individual standard deviation across items, and the number of distinct response options used. A composite flag combined these criteria.

\emph{Acquiescence modelling.} Two complementary treatments were applied. A random-intercept common factor was specified with all 67 loadings fixed to unity and orthogonal to the substantive factors \cite{maydeu2006randomintercept}. As a robust check, each item was residualised on the agreement index of \eqref{eq:ars} through ordinary least squares, and construct correlations were recomputed. Within-person centring (ipsatization), $\tilde{x}_{ij}=x_{ij}-a_i$, was used strictly as a diagnostic, its known negative-dependency artefact interpreted through pattern rather than magnitude.

\emph{Criterion validity by known groups.} Because ordinal exposure variables are comparatively resistant to a uniform agreement style, composite scores were compared across exposure levels. One-way analysis of variance produced
\begin{equation}
F=\frac{\text{MS}_{\text{between}}}{\text{MS}_{\text{within}}},\qquad
\eta^2=\frac{\text{SS}_{\text{between}}}{\text{SS}_{\text{total}}},
\label{eq:anova}
\end{equation}
with $\eta^2$ benchmarked at 0.01, 0.06, and 0.14 for small, medium, and large effects \cite{cohen1988power}. The nonparametric Kruskal-Wallis $H$ confirmed group differences without normality assumptions, and Spearman's $\rho$ quantified the monotone trend across ordered exposure levels. Bias-corrected bootstrap confidence intervals for $\eta^2$ were obtained from 2{,}000 resamples.

\subsection{Proposed Analytical Framework}
\begin{figure}[!t]
\centering
\begin{tikzpicture}[
  font=\scriptsize,
  >={Latex[length=1.5mm]},
  box/.style={rectangle, rounded corners=2.5pt, draw=black!45, line width=0.5pt,
              minimum width=1.7cm, minimum height=0.82cm, align=center, inner sep=2pt},
  expo/.style={box, fill=red!7,   draw=red!45},
  mech/.style={box, fill=orange!10, draw=orange!55},
  outc/.style={box, fill=teal!10,  draw=teal!55},
  modr/.style={box, fill=black!5,  draw=black!45, minimum width=1.5cm, minimum height=0.66cm},
  path/.style={->, line width=0.5pt, draw=black!55},
  modpath/.style={->, line width=0.5pt, draw=black!45, dashed},
  siglab/.style={circle, fill=white, inner sep=0.4pt, font=\scriptsize}
]

% ---- tier band backgrounds (subtle grouping) ----
\begin{scope}[on background layer]
  \fill[red!3]   (-1.25,-1.75) rectangle (0.95,1.7);
  \fill[orange!3] (2.05,-1.75) rectangle (4.25,1.7);
  \fill[teal!3]  (5.35,-1.75) rectangle (7.55,1.7);
\end{scope}

% ---- tier captions ----
\node[font=\scriptsize\itshape, text=red!55]   at (-0.15,2.05) {Exposure};
\node[font=\scriptsize\itshape, text=orange!65] at (3.15,2.05)  {Mechanism};
\node[font=\scriptsize\itshape, text=teal!60]   at (6.45,2.05)  {Outcome};

% ---- nodes ----
\node[expo] (sfca) at (-0.15,0.7)  {Short-form\\addiction};
\node[expo] (no)   at (-0.15,-1.0) {Notification\\overload};
\node[mech] (af)   at (3.15,0.7)   {Attention\\fragmentation};
\node[mech] (cf)   at (3.15,-1.0)  {Cognitive\\fatigue};
\node[outc] (dle)  at (6.45,0.7)   {Deep learning\\engagement};
\node[outc] (ab)   at (6.45,-1.0)  {Academic\\burnout};
\node[modr] (sr)   at (1.5,1.55)   {Self-\\regulation};

% moderation junction on the addiction->fragmentation path
\coordinate (j) at (1.5,0.7);

% ---- antecedent -> mechanism ----
\draw[path] (sfca) -- (af);
\draw[path] (sfca.south) |- (cf.west);
\draw[path] (no.north) |- (af.west);
\draw[path] (no) -- (cf);

% ---- mechanism -> outcome (orthogonal, no diagonal collisions) ----
\draw[path] (af) -- node[siglab,pos=0.5]{$-$} (dle);
\draw[path] (cf) -- node[siglab,pos=0.5]{$+$} (ab);
\draw[path] (af.south) |- ($(ab.west)+(0,0.16)$) node[siglab,pos=0.94]{$+$};
\draw[path] (cf.north) |- ($(dle.west)+(0,-0.16)$) node[siglab,pos=0.94]{$-$};

% ---- moderator: self-regulation acts on the addiction->fragmentation link ----
\draw[modpath] (sr.south) -- (j);
\fill[black!45] (j) circle (0.9pt);

\end{tikzpicture}
\caption{A priori theoretical framework. Exposure constructs (left) are hypothesised to raise attention fragmentation and cognitive fatigue (centre), which in turn erode deep learning engagement and elevate academic burnout (right). Self-regulation (dashed) moderates the addiction-to-fragmentation pathway, indicated by the junction on that link. Signed labels denote the predicted direction of each effect. The analysis tests, rather than assumes, the measurement prerequisites of this model.}
\label{fig:framework}
\end{figure}
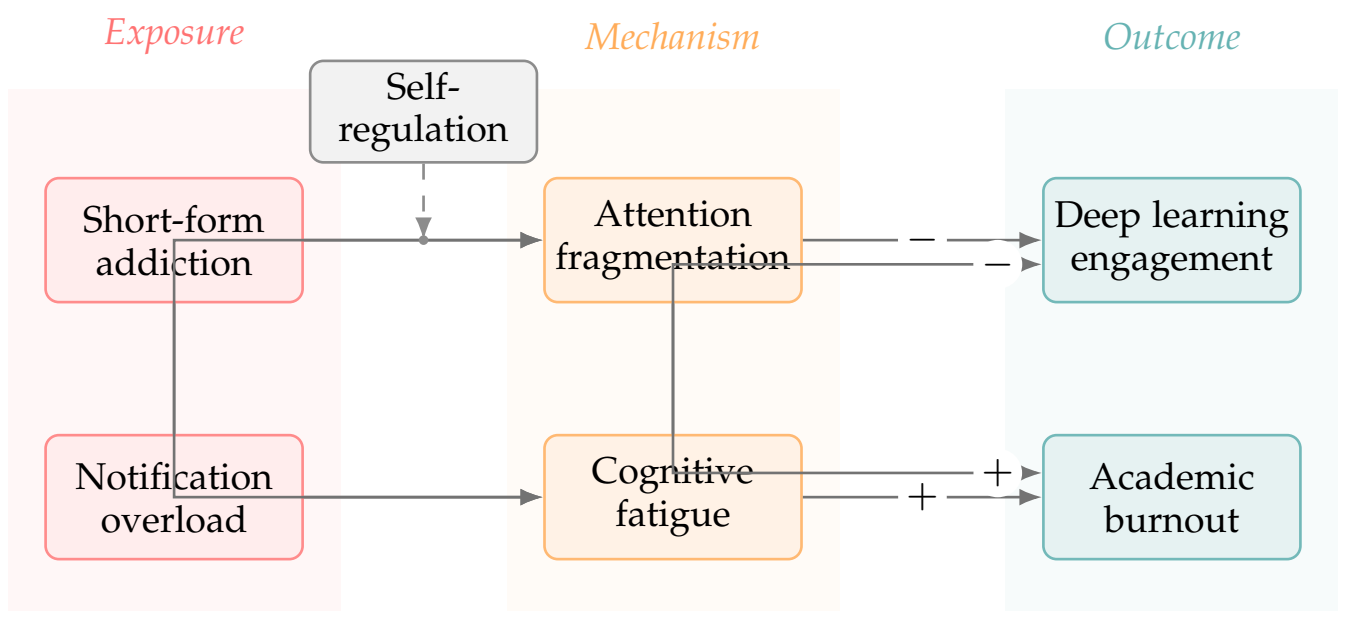

Figure~\ref{fig:framework} depicts the a priori theoretical model that motivated the instrument: exposure and its psychological correlates drive attention fragmentation and cognitive fatigue, which in turn erode deep learning engagement and inflate academic burnout, with self-regulation buffering the initial pathway. The analytic protocol does not presuppose this model; it tests each measurement prerequisite the model requires and, where those prerequisites fail, redirects inference toward the components the data can support.

\subsection{Experimental Setup}
All coding and analyses were implemented in Google Colaboratory using Python 3, with the hardware accelerator set to CPU. Data handling and construct scoring relied on \texttt{pandas} and \texttt{NumPy}, with the item response matrix ingested through \texttt{pandas.read\_csv}, column labels canonicalised by regular-expression matching, and composite and agreement indices computed through vectorised \texttt{DataFrame.mean} operations following Equations~\eqref{eq:composite} and \eqref{eq:ars}. Distributional screening used \texttt{scipy.stats.skew} and \texttt{scipy.stats.kurtosis} with bias correction, evaluated against the severity thresholds of Section~III.

Reliability and latent-structure procedures were built on \texttt{factor\_analyzer}. Cronbach's alpha and McDonald's omega were computed from a single-factor solution obtained with \texttt{FactorAnalyzer} under the minimum-residual method, and corrected item-total correlations were derived directly from the correlation matrix. Factorability was assessed with \texttt{calculate\_kmo} and \texttt{calculate\_bartlett\_sphericity}, the number of retainable factors was determined by a custom parallel-analysis routine that compared observed eigenvalues from \texttt{numpy.linalg.eigvalsh} against the ninety-fifth percentile of eigenvalues simulated from random normal data, and the theorised facet structure was probed by exploratory factor analysis under promax rotation. Average variance extracted and composite reliability (Equation~\eqref{eq:ave}) were computed from the standardised loadings, and a guided item-retention loop iteratively removed the weakest indicator until the validity thresholds were met or a three-item floor was reached.

Response-quality and response-style diagnostics were implemented directly in \texttt{NumPy} and \texttt{pandas}. Careless responding was screened through a longstring routine, intra-individual dispersion via \texttt{DataFrame.std}, and response-option range via \texttt{Series.nunique}. Acquiescence was modelled in two ways: a random-intercept common factor with unit loadings fixed across all items and estimated with the \texttt{semopy} structural-equation package, and a robust regression-partialling of the agreement index solved in closed form with \texttt{NumPy}. Criterion validity by known groups used \texttt{scipy.stats.f\_oneway} and \texttt{scipy.stats.kruskal} for the omnibus tests, a Spearman correlation via \texttt{pandas.Series.corr} for the monotone trend, and a custom bootstrap over 2{,}000 resamples for the confidence intervals on $\eta^2$.

%=====================================================================
\section{Results}
\begin{figure}[!t]
\centering
\includegraphics[width=\columnwidth]{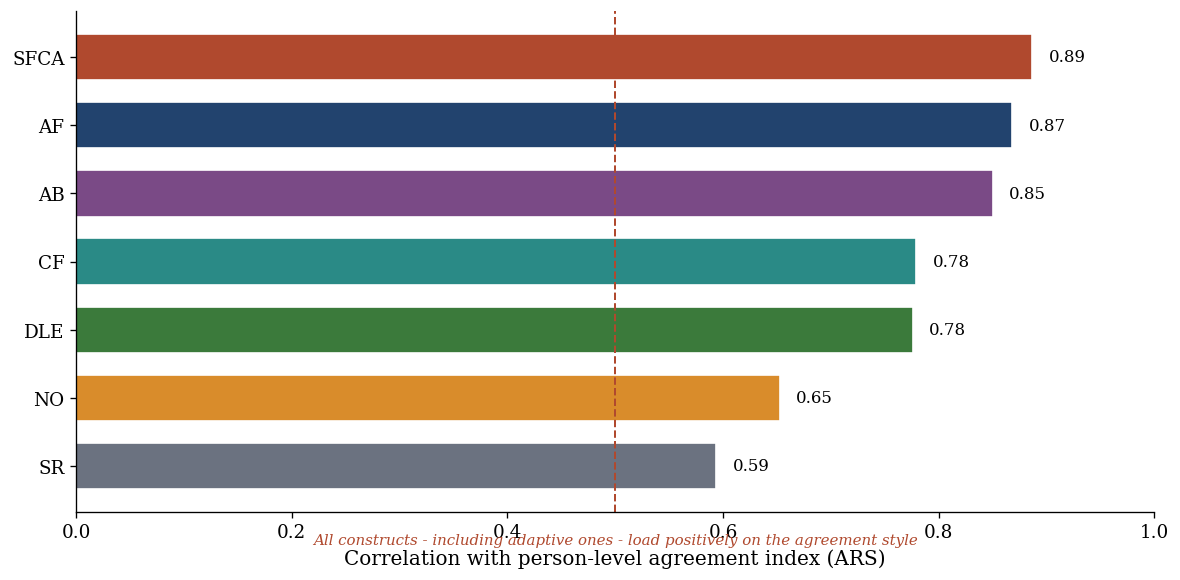}
\caption{Correlation of each construct composite with the person-level agreement index. All constructs, including the adaptive ones (deep learning engagement and self-regulation), load positively and substantially, evidencing a pervasive uniform agreement style.}
\label{fig:acq}
\end{figure}

%=====================================================================
The results are organised to mirror the analytic protocol, moving from sample description and distributional properties through the measurement audit to the response-style diagnosis and, finally, the criterion-validity evidence. Table~\ref{tab:sample} profiles the sample; Table~\ref{tab:reliability} reports reliability and dimensionality; Table~\ref{tab:validity} reports convergent validity; Table~\ref{tab:acq} reports the acquiescence diagnostics; Table~\ref{tab:corr} reports the sign audit of composite correlations; and Tables~\ref{tab:daily} and \ref{tab:check} report the dose-response evidence. Figures~\ref{fig:dose} through \ref{fig:acq} visualise the central findings.

\subsection{Sample and Distributional Properties}
The realised sample was demographically balanced on gender and concentrated in the mid-undergraduate to early-postgraduate range, as summarised in Table~\ref{tab:sample}. Item distributions were well behaved: skewness ranged from $-0.172$ to $0.648$ and excess kurtosis from $-0.902$ to $-0.390$, with no item breaching conventional severity thresholds. The mild, symmetric non-normality typical of Likert data justifies estimators that make no multivariate-normality assumption and rules out extreme distributional artefacts as an explanation for later findings. Construct composite means clustered tightly between 2.59 and 2.75 on the five-point scale, a homogeneity that foreshadows the limited discriminating variance recovered in subsequent stages.

\begin{table}[!t]
\caption{Sample characteristics ($N=863$).}
\label{tab:sample}
\centering
\footnotesize
\begin{tabular}{@{}llrr@{}}
\toprule
Variable & Category & $n$ & \% \\
\midrule
\multirow{4}{*}{Age} & 16 to 20 & 117 & 13.6 \\
 & 21 to 23 & 348 & 40.3 \\
 & 24 to 26 & 348 & 40.3 \\
 & Above 26 & 50 & 5.8 \\
\midrule
\multirow{2}{*}{Gender} & Male & 433 & 49.9 \\
 & Female & 434 & 50.1 \\
\midrule
\multirow{3}{*}{University type} & Public & 151 & 17.5 \\
 & Private & 634 & 73.5 \\
 & Other & 78 & 9.0 \\
\midrule
\multirow{4}{*}{Academic level} & Undergraduate & 347 & 40.2 \\
 & Master's & 297 & 34.4 \\
 & MPhil & 188 & 21.8 \\
 & PhD & 31 & 3.6 \\
\midrule
\multirow{5}{*}{Daily time} & Less than 30 min & 80 & 9.3 \\
 & 30 min to 1 h & 218 & 25.3 \\
 & 1 to 2 h & 295 & 34.2 \\
 & 2 to 4 h & 195 & 22.6 \\
 & More than 4 h & 75 & 8.7 \\
\midrule
\multirow{6}{*}{Checks in study} & Never & 142 & 16.5 \\
 & Rarely & 188 & 21.8 \\
 & Sometimes & 312 & 36.2 \\
 & Often & 136 & 15.8 \\
 & Very often & 52 & 6.0 \\
 & Always & 33 & 3.8 \\
\bottomrule
\end{tabular}
\end{table}

\subsection{Reliability and Dimensionality}
Pooled internal consistency appeared strong at the construct level, with Cronbach's alpha of 0.841 for addiction, 0.811 for attention fragmentation, 0.830 for deep learning engagement, and 0.814 for academic burnout, and McDonald's omega tracking these values closely (Table~\ref{tab:reliability}). This apparent strength, however, was contradicted at the facet level, where first-order subscale reliabilities fell between 0.529 and 0.696, and mean inter-item correlations hovered near 0.25 both within facets and across whole constructs. That the four items of a facet cohered barely more than sixteen items drawn from an entire construct is the first signal that the theorised structure was not operating as designed; the high pooled alpha reflects item count rather than tight, distinct facets.

Dimensionality assessment confirmed the diagnosis decisively. Sampling adequacy was excellent, with Kaiser-Meyer-Olkin indices between 0.898 and 0.922 and Bartlett's test significant for every construct, so the correlation matrices were unambiguously factorable. Yet parallel analysis retained exactly one factor for each of the four multidimensional constructs: the first eigenvalue was large (4.2 to 4.7) while the second already fell below its random-data threshold. Forced oblique extraction to the theorised number of factors failed to reproduce the assigned facets, with items loading diffusely across factors. The four constructs are therefore empirically unidimensional, and the higher-order specification the instrument presumed is not supported.

\begin{table}[!t]
\caption{Reliability and dimensionality by construct ($N=863$). PA denotes factors retained by parallel analysis against the theorised number.}
\label{tab:reliability}
\centering
\footnotesize
\begin{tabular}{@{}lrrrrr@{}}
\toprule
Construct & $k$ & $\alpha$ & $\omega$ & KMO & PA / theory \\
\midrule
SFCA & 16 & 0.841 & 0.841 & 0.922 & 1 / 4 \\
AF   & 12 & 0.811 & 0.811 & 0.898 & 1 / 3 \\
DLE  & 12 & 0.830 & 0.830 & 0.910 & 1 / 3 \\
AB   & 12 & 0.814 & 0.814 & 0.902 & 1 / 3 \\
NO   & 5  & 0.654 & 0.655 & --    & 1 / 1 \\
CF   & 5  & 0.681 & 0.686 & --    & 1 / 1 \\
SR   & 5  & 0.651 & 0.654 & --    & 1 / 1 \\
\bottomrule
\end{tabular}
\end{table}

\subsection{Convergent Validity}
Treated as unidimensional, no construct achieved convergent validity. Average variance extracted ranged from 0.250 to 0.307 across the full item sets, uniformly below the 0.50 criterion, with standardised loadings between 0.40 and 0.64 (Table~\ref{tab:validity}). Because average variance extracted is the mean squared loading, a ceiling item correlating near 0.64 with its factor cannot lift the aggregate to threshold, and the guided retention loop confirmed this: even after trimming each construct to its three strongest indicators, the best attainable value was 0.385, still short of 0.50, while composite reliability fell as items were removed. Item deletion therefore cannot rescue the model, because the deficiency is distributed across all indicators rather than localised in a few weak ones.

\begin{table}[!t]
\caption{Convergent validity (full item sets). All constructs fail the $\text{AVE}\geq0.50$ criterion.}
\label{tab:validity}
\centering
\footnotesize
\begin{tabular}{@{}lrrrr@{}}
\toprule
Construct & $k$ & AVE & $\rho_c$ & Loading range \\
\midrule
SFCA & 16 & 0.250 & 0.841 & 0.40 to 0.58 \\
AF   & 12 & 0.264 & 0.811 & 0.42 to 0.56 \\
NO   & 5  & 0.277 & 0.655 & 0.47 to 0.60 \\
CF   & 5  & 0.307 & 0.686 & 0.43 to 0.61 \\
DLE  & 12 & 0.292 & 0.830 & 0.45 to 0.64 \\
AB   & 12 & 0.269 & 0.814 & 0.47 to 0.57 \\
SR   & 5  & 0.277 & 0.654 & 0.40 to 0.58 \\
\bottomrule
\end{tabular}
\end{table}

\subsection{Careless Responding is Not the Cause}
A natural hypothesis for uniformly weak factor loadings is inattentive responding, and this was tested directly. A composite flag combining longstring, intra-individual variability, and response-option range identified 8.0\% of respondents (69 cases). Removing them, however, lowered rather than raised average variance extracted for every construct, with reductions of approximately 0.06 in each case. This inversion is the opposite of the careless-responding signature: the flagged respondents were disproportionately uniform high endorsers whose removal further attenuated the already weak inter-item structure. The measurement weakness is therefore intrinsic to how the sample engaged with the instrument, not an artefact of a deletable subset.

\subsection{Acquiescence as the Operative Mechanism}
\begin{figure}[!t]
\centering
\includegraphics[width=\columnwidth]{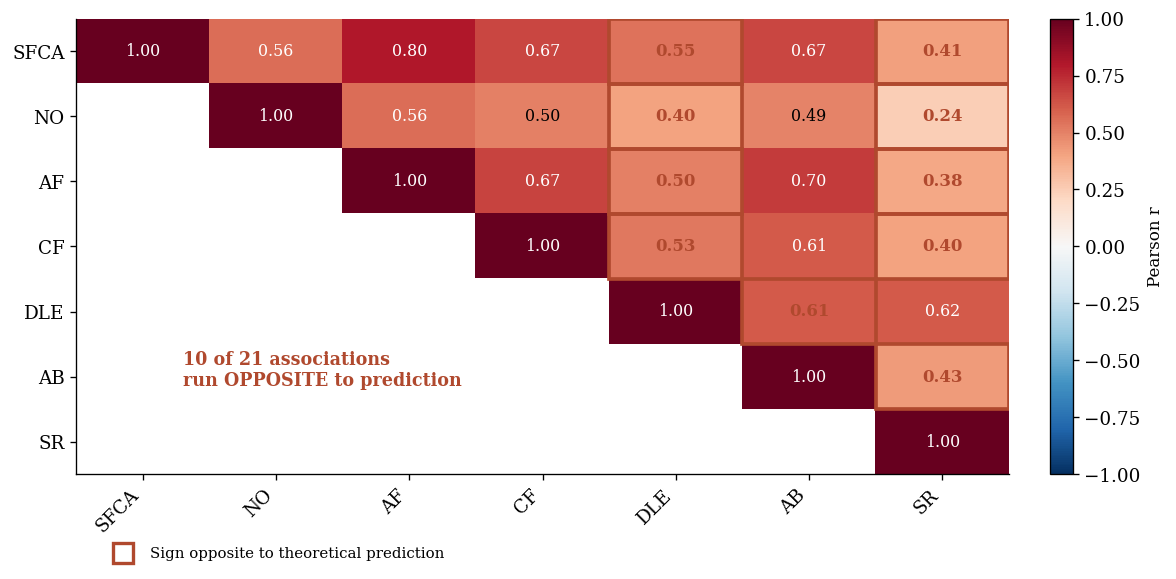}
\caption{Composite correlations among the seven constructs. Cells outlined in red, with bold red values, carry a sign opposite to theoretical prediction; the affected pairs are precisely the maladaptive-versus-adaptive associations an agreement response style is expected to invert.}
\label{fig:corr}
\end{figure}

The decisive evidence concerns response style. The person-level agreement index of \eqref{eq:ars} correlated strongly and positively with every construct, from 0.594 for self-regulation to 0.887 for addiction (Table~\ref{tab:acq}), implying that general endorsement explained the majority of variance in each scale, including the two adaptive constructs that theory predicts should move opposite to the rest. Consistent with this, the mean of the maladaptive composites correlated $+0.623$ with the mean of the adaptive composites, precisely the wrong sign for a model in which addiction and burnout should accompany lower deep learning and weaker self-regulation. The correlation heatmap in Figure~\ref{fig:corr} makes the scope explicit: 10 of 21 theory-relevant associations carry a sign opposite to prediction, and all ten fall in the maladaptive-versus-adaptive pairings where a uniform agreement style is expected to invert the relationship.

Removing the agreement component restored the theorised structure for exactly those pairings. Partialling the agreement index reversed the addiction-to-deep-learning correlation from $+0.547$ to $-0.486$, the fragmentation-to-deep-learning correlation from $+0.503$ to $-0.547$, the addiction-to-self-regulation correlation from $+0.411$ to $-0.312$, and the deep-learning-to-burnout correlation from $+0.614$ to $-0.138$ (Table~\ref{tab:acq}). The uniformity of this sign recovery, confined to the pairings the psychometric literature predicts \cite{primi2020acquiescence}, identifies acquiescence rather than absent theory as the source of the anomaly. The correction is diagnostic, not curative, however: it simultaneously drove same-pole maladaptive correlations toward zero and collapsed convergent validity further, because a single agreement predictor so collinear with every construct removes content variance along with style variance. The random-intercept specification, constrained orthogonal to the substantive factors and fixed at unit loadings, left construct correlations essentially unchanged, since the uniformly keyed substantive factors reabsorbed the agreement variance. Neither treatment yields a defensible latent model, which is itself the finding: the battery cannot support the intended structural analysis.

\begin{table}[!t]
\caption{Acquiescence diagnostics. Left: correlation of each composite with the agreement index. Right: raw versus agreement-partialled correlations for the key theory-relevant pairs (all predicted negative).}
\label{tab:acq}
\centering
\footnotesize
\begin{tabular}{@{}lr@{\hskip 10pt}lrr@{}}
\toprule
Construct & $r$ with ARS & Pair & Raw & Partialled \\
\midrule
SFCA & 0.887 & SFCA--DLE & $+0.547$ & $-0.486$ \\
AF   & 0.869 & AF--DLE   & $+0.503$ & $-0.547$ \\
AB   & 0.850 & CF--DLE   & $+0.526$ & $-0.199$ \\
CF   & 0.780 & SFCA--SR  & $+0.411$ & $-0.312$ \\
DLE  & 0.776 & AF--SR    & $+0.383$ & $-0.333$ \\
NO   & 0.653 & DLE--AB   & $+0.614$ & $-0.138$ \\
SR   & 0.594 & AB--SR    & $+0.429$ & $-0.179$ \\
\bottomrule
\end{tabular}
\end{table}

\begin{table}[!t]
\caption{Composite inter-construct correlations. Entries in \rc{red} carry a sign opposite to theoretical prediction (10 of 21 theory-relevant pairs).}
\label{tab:corr}
\centering
\footnotesize
\setlength{\tabcolsep}{4.2pt}
\begin{tabular}{@{}lrrrrrrr@{}}
\toprule
 & SFCA & NO & AF & CF & DLE & AB & SR \\
\midrule
SFCA & 1.00 & .56 & .80 & .67 & \rc{.55} & .67 & \rc{.41} \\
NO   &      & 1.00 & .56 & .50 & \rc{.40} & .49 & \rc{.24} \\
AF   &      &      & 1.00 & .67 & \rc{.50} & .70 & \rc{.38} \\
CF   &      &      &      & 1.00 & \rc{.53} & .61 & \rc{.40} \\
DLE  &      &      &      &      & 1.00 & \rc{.61} & .62 \\
AB   &      &      &      &      &      & 1.00 & \rc{.43} \\
SR   &      &      &      &      &      &      & 1.00 \\
\bottomrule
\end{tabular}
\end{table}

\subsection{Dose-Response Criterion Validity}
\begin{figure}[!t]
\centering
\includegraphics[width=\columnwidth]{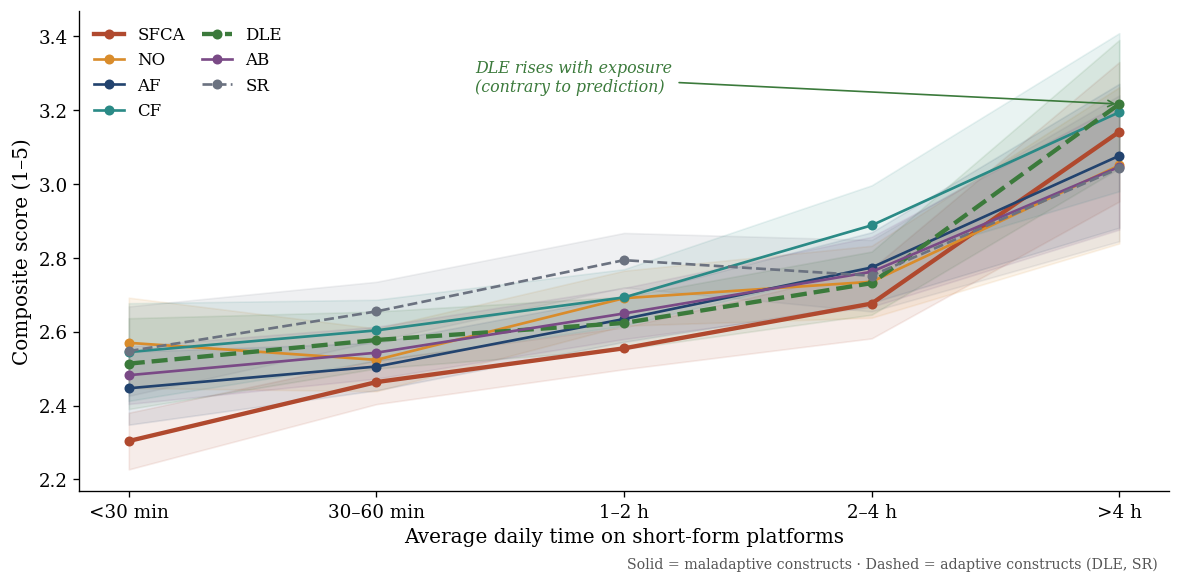}
\caption{Composite means across five levels of daily short-form platform time, with 95\% confidence bands. Maladaptive constructs (solid) rise monotonically with exposure; the deep learning engagement line (dashed) also rises, contrary to prediction, reflecting residual agreement variance.}
\label{fig:dose}
\end{figure}

\begin{figure}[!t]
\centering
\includegraphics[width=\columnwidth]{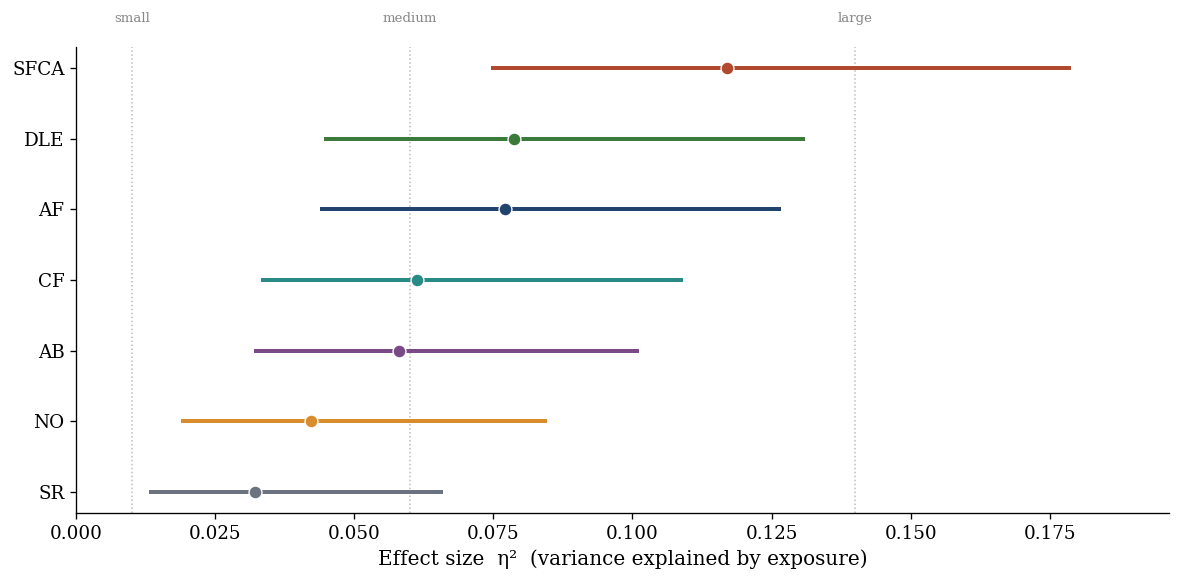}
\caption{Effect size ($\eta^2$) of daily-time exposure on each construct, with bias-corrected bootstrap 95\% confidence intervals. Dotted lines mark small (0.01), medium (0.06), and large (0.14) benchmarks. Short-form content addiction shows the strongest exposure effect.}
\label{fig:eta}
\end{figure}

Where the self-report correlations were compromised, the behavioural exposure comparisons were not. Composite scores rose monotonically across levels of daily platform time for every maladaptive construct, and each gradient was significant by both analysis of variance and the nonparametric Kruskal-Wallis test (Table~\ref{tab:daily}, Figure~\ref{fig:dose}). Addiction showed the largest effect, climbing from 2.30 among the lightest users to 3.14 among the heaviest ($F(4,858)=28.43$, $p<0.001$, $\eta^2=0.117$, $\rho=0.273$), with attention fragmentation ($\eta^2=0.077$), cognitive fatigue ($\eta^2=0.061$), academic burnout ($\eta^2=0.058$), and notification overload ($\eta^2=0.042$) following the same ordered pattern. Because group-mean contrasts do not depend on the item-level correlations that acquiescence distorts, this dose-response structure constitutes genuine, response-style-robust criterion validity. The effect-size ordering, with the most face-valid construct strongest, is exactly what a substantive exposure-strain relationship predicts (Figure~\ref{fig:eta}).

The two adaptive constructs behaved anomalously in a manner that corroborates the acquiescence account rather than contradicting the substantive story. Deep learning engagement rose with exposure ($\eta^2=0.079$) where theory predicts a decline, and self-regulation showed a weak positive trend that was significant for daily time but not for in-study checking ($F(5,857)=1.50$, $p=0.19$; Table~\ref{tab:check}). Heavier users who endorse the maladaptive items more strongly also endorse the adaptive items more strongly, so the residual agreement style surfaces as a paradoxical positive gradient on precisely the two constructs that should oppose the rest. The dose-response evidence and the acquiescence diagnosis thus form a single coherent picture rather than two competing ones.

\begin{table}[!t]
\caption{Dose-response by average daily time (five ordered levels; $\text{df}=4,858$). Kruskal-Wallis $p<0.001$ for all constructs. Expected trend positive for maladaptive constructs, negative for adaptive (DLE, SR).}
\label{tab:daily}
\centering
\footnotesize
\begin{tabular}{@{}lrrrl@{}}
\toprule
Construct & $F$ & $\eta^2$ & $\rho$ & Trend \\
\midrule
SFCA & 28.43 & 0.117 & 0.273 & as predicted \\
AF   & 17.93 & 0.077 & 0.218 & as predicted \\
CF   & 14.02 & 0.061 & 0.208 & as predicted \\
AB   & 13.20 & 0.058 & 0.205 & as predicted \\
NO   &  9.46 & 0.042 & 0.160 & as predicted \\
DLE  & 18.32 & 0.079 & 0.218 & \rc{opposite} \\
SR   &  7.11 & 0.032 & 0.141 & \rc{opposite} \\
\bottomrule
\end{tabular}
\end{table}

\begin{table}[!t]
\caption{Dose-response by in-study checking frequency (six ordered levels; $\text{df}=5,857$).}
\label{tab:check}
\centering
\footnotesize
\begin{tabular}{@{}lrrrl@{}}
\toprule
Construct & $F$ & $\eta^2$ & $p$ & Trend \\
\midrule
SFCA & 19.71 & 0.103 & $<0.001$ & as predicted \\
AF   & 17.21 & 0.091 & $<0.001$ & as predicted \\
AB   & 13.39 & 0.072 & $<0.001$ & as predicted \\
CF   & 11.49 & 0.063 & $<0.001$ & as predicted \\
NO   &  6.94 & 0.039 & $<0.001$ & as predicted \\
DLE  &  7.03 & 0.039 & $<0.001$ & \rc{opposite} \\
SR   &  1.50 & 0.009 & 0.19     & n.s. \\
\bottomrule
\end{tabular}
\end{table}

%=====================================================================
\section{Discussion}
%=====================================================================
\subsection{Why the Results Occurred}
The joint pattern of a robust behavioural dose-response and a collapsed self-report measurement model is not contradictory once acquiescence is recognised as the connecting mechanism. A uniform agreement style adds a positive constant to a respondent's answers across every item, which mechanically inflates same-pole correlations, deflates opposite-pole correlations, and props up internal consistency through item count while leaving little content-specific variance to define distinct factors \cite{primi2020acquiescence,baumgartner2001response}. This single mechanism accounts for each stage of the audit: the high pooled alpha beside weak facet reliabilities, the one-factor parallel-analysis solutions, the sub-threshold average variance extracted, the paradoxical worsening of validity when uniform endorsers were removed, and the wrong-signed correlations that reverted to theory once the agreement component was partialled. The behavioural comparisons survived because group-mean contrasts on ordinal exposure do not rely on the inter-item covariances that the style corrupts, so the substantive gradient remained visible even as the latent structure dissolved.

\subsection{Comparison with Previous Studies}
The substantive gradient aligns with the direction reported across recent short-form video research, which consistently associates heavier or more compulsive use with greater academic anxiety, disengagement, cognitive failure, and attentional difficulty \cite{li2024shortform,haliti2024reels,shuja2025mindfulness,singh2025attention}. Where the present work diverges is in the epistemic status it assigns to that convergence. Much of the prior evidence is drawn from single-source, same-direction Likert batteries analysed by latent models whose discriminant validity is asserted rather than demonstrated, and the current audit shows that such designs can satisfy model-fit criteria on the strength of shared method variance. The finding that partialling a single agreement factor reverses the sign of the addiction-to-deep-learning relationship is a concrete instance of the distortion the psychometric literature has long warned about \cite{maydeu2006randomintercept,martins2024comparison}, and it suggests that some published coefficients in this domain may be partly stylistic. The contribution is thus corrective as much as confirmatory: it validates the field's directional claims through exposure comparisons while cautioning against over-reliance on its correlational machinery.

\subsection{Theoretical Implications}
Theoretically, the exposure gradients are consistent with the integrative frame that motivated the study. That addiction, fragmentation, fatigue, and burnout all intensify with exposure is compatible with a limited-capacity account in which continual high-arousal switching taxes attentional resources \cite{lang2000limited}, and with a resource-depletion account in which repeated taxation without recovery culminates in burnout \cite{hobfoll1989cor}. The measurement findings, however, sharpen a methodological corollary of these theories: constructs that are conceptually adjacent and measured with uniformly valenced items are especially prone to stylistic conflation, so theoretical models that presume clean separation among such constructs require instrumentation designed to deliver it. The study therefore refines rather than refutes the guiding theory, relocating the burden of proof from the plausibility of the model to the adequacy of its measurement.

\subsection{Practical Implications}
Practically, the results carry implications for researchers, educators, and instrument designers. For researchers, the immediate lesson is that behavioural-addiction batteries intended for latent modelling must be balanced with reverse-keyed items, fitted with attention checks and forced response variation, and accompanied by explicit acquiescence diagnostics and, where warranted, random-intercept correction \cite{maydeu2006randomintercept,savalei2014robustness}. For educators and institutions, the dose-response evidence, modest in effect size yet consistent and significant, supports attention-aware pedagogy, notification-hygiene guidance, and self-regulation support targeted at heavier users, who show the steepest strain profiles. For instrument designers in this rapidly growing area, the study offers a cautionary template: a scale can post enviable reliability coefficients and still lack the convergent and discriminant validity that any structural inference requires.

%=====================================================================
\section{Conclusion}
%=====================================================================
This investigation set out to characterise the relationship between short-form content exposure and cognitive-academic strain and to audit the measurement instrument on which such characterisations typically depend. Two findings resulted. Self-reported exposure exhibits a coherent, monotone dose-response relationship with short-form content addiction, attention fragmentation, notification overload, cognitive fatigue, and academic burnout, robust across parametric and nonparametric tests and strongest for the most face-valid construct. Simultaneously, the accompanying seven-construct self-report battery fails its measurement prerequisites, and a pervasive acquiescence style is shown to be responsible, one that masks and reverses the theoretically negative relationships central to the field.

The contributions follow from this pairing. The study delivers a response-style-robust confirmation of the exposure-strain gradient that does not rest on contaminated inter-item correlations, and it provides a transparent, reproducible demonstration that a uniformly worded battery can present strong reliability while failing validity, with acquiescence as the diagnosable cause. In doing so it connects an applied literature that rarely tests for response style with a psychometric literature that rarely examines the instruments now proliferating in that domain.

Several limitations bound these conclusions. The design is cross-sectional and single-source, so causal and temporal claims are not warranted; the sample skews toward younger, predominantly private-university students; and the exposure measures are themselves self-reported categories rather than objective usage logs. The instrument's uniform keying, the very feature that produced the acquiescence problem, also prevented a clean model-based correction. Future work should recollect with a balanced, shortened instrument that mixes item valence, embeds attention checks, and where possible incorporates device-logged usage, then re-estimate the moderated-mediation model that the present data could not sustain. Longitudinal and experience-sampling designs would further separate exposure from outcome in time. Pursued along these lines, the substantive gradient documented here can be tested as a mechanism rather than merely observed as an association.

%=====================================================================
\section*{Ethical Approval}
The study was conducted in accordance with the ethical principles of the Declaration of Helsinki and recognised standards for research involving human participants. Ethical approval was obtained from the Ethical Approval Committee of Skill Morph Research Lab., Skill Morph, Dhaka, Bangladesh (Approval No. SkillMorph/ES/2026/01(7)). Participation was voluntary and anonymous. Electronic informed consent was obtained from all participants through an opening consent statement prior to questionnaire completion. No personally identifiable information was collected. 

\section*{Funding Statement}
This research received no external funding.

\section*{Conflict of Interest}
The author declares no conflict of interest.

\section*{Author Contributions}
Md.~Alamgir~Hossain is the sole author and was solely responsible for conceptualisation, study design, instrument development, data collection, analysis, interpretation, and manuscript preparation.

\section*{Data Availability}
The dataset and analysis code are available from the author on reasonable request.

\section*{Acknowledgements}
The authors gratefully acknowledge the support of Skill Morph Research Lab., Skill Morph, Dhaka, Bangladesh.

%=====================================================================


\begin{thebibliography}{99}
\itemsep=0pt

\bibitem{roberts2025technology}
J.~A.~Roberts and M.~E.~David, ``Technology affordances, social media engagement, and social media addiction: An investigation of TikTok, Instagram Reels, and YouTube Shorts,'' \emph{Cyberpsychology, Behavior, and Social Networking}, vol.~28, no.~5, pp.~318--325, May 2025, doi: 10.1089/cyber.2024.0338.

\bibitem{yin2024virality}
H.~Yin, ``From virality to engagement: Examining the transformative impact of social media, short video platforms, and live streaming on information dissemination and audience behavior in the digital age,'' \emph{Advances in Social Behavior Research}, vol.~14, pp.~10--14, 2024.

\bibitem{hsieh2025smartphone}
C.-Y.~Hsieh, ``The impact of smartphone usage frequency on university students' academic performance: A meta-analysis of moderating factors,'' \emph{Acta Psychologica}, vol.~259, Art. no.~105374, 2025, doi: 10.1016/j.actpsy.2025.105374.

\bibitem{firth2024online}
J.~Firth, J.~Torous, J.~F.~López-Gil, J.~Linardon, A.~Milton, J.~Lambert, L.~Smith, I.~Jarić, H.~Fabian, D.~Vancampfort, H.~Onyeaka, F.~B.~Schuch, and J.~A.~Firth, ``From `online brains' to `online lives': Understanding the individualized impacts of Internet use across psychological, cognitive and social dimensions,'' \emph{World Psychiatry}, vol.~23, no.~2, pp.~176--190, Jun. 2024, doi: 10.1002/wps.21188.

\bibitem{clemente2024digital}
V.~J.~Clemente-Suárez, A.~I.~Beltrán-Velasco, S.~Herrero-Roldán, S.~Rodriguez-Besteiro, I.~Martínez-Guardado, A.~Martín-Rodríguez, and J.~F.~Tornero-Aguilera, ``Digital device usage and childhood cognitive development: Exploring effects on cognitive abilities,'' \emph{Children}, vol.~11, no.~11, Art. no.~1299, Oct. 2024, doi: 10.3390/children11111299.

\bibitem{li2024shortform}
G.~Li, Y.~Geng, and T.~Wu, ``Effects of short-form video app addiction on academic anxiety and academic engagement: The mediating role of mindfulness,'' \emph{Frontiers in Psychology}, vol.~15, art.~1428813, 2024, doi: 10.3389/fpsyg.2024.1428813.

\bibitem{haliti2024reels}
T.~Haliti-Sylaj and A.~Sadiku, ``Impact of short reels on attention span and academic performance of undergraduate students,'' \emph{Eurasian Journal of Applied Linguistics}, vol.~10, no.~3, pp.~60--68, 2024, doi: 10.32601/ejal.10306.

\bibitem{shuja2025mindfulness}
T.~Shuja, Z.~Ahmad, and W.~Wazir, ``Short-form video addiction, mindfulness, and cognitive failure among undergraduate university students,'' \emph{Journal of Nursing and Allied Health}, vol.~3, no.~3, pp.~160--166, 2025.

\bibitem{baumgartner2001response}
H.~Baumgartner and J.~E.~M.~Steenkamp, ``Response styles in marketing research: A cross-national investigation,'' \emph{Journal of Marketing Research}, vol.~38, no.~2, pp.~143--156, 2001, doi: 10.1509/jmkr.38.2.143.18840.

\bibitem{primi2020acquiescence}
R.~Primi, D.~Santos, O.~P.~John, and F.~De~Fruyt, ``Comparison of classical and modern methods for measuring and correcting for acquiescence,'' \emph{European Journal of Psychological Assessment}, vol.~36, no.~1, pp.~1--12, 2020, doi: 10.1111/bmsp.12168.

\bibitem{maydeu2006randomintercept}
A.~Maydeu-Olivares and D.~L.~Coffman, ``Random intercept item factor analysis,'' \emph{Psychological Methods}, vol.~11, no.~4, pp.~344--362, Dec. 2006, doi: 10.1037/1082-989X.11.4.344.

\bibitem{martins2024comparison}
G.~H.~Martins, F.~Valentini, B.~Bonf\'a-Araujo, A.~S.~Ferraz, F.~Otoni, and A.~P.~Gon\c{c}alves, ``Comparison of methods for controlling acquiescence bias in balanced and unbalanced scales,'' \emph{Psico-USF}, vol.~28, no.~4, pp.~651--667, 2024, doi: 10.1590/1413-82712023280401.

\bibitem{savalei2014robustness}
V.~Savalei and C.~F.~Falk, ``Recovering substantive factor loadings in the presence of acquiescence bias: A comparison of three approaches,'' \emph{Multivariate Behavioral Research}, vol.~49, no.~5, pp.~407--424, Sep.–Oct. 2014, doi: 10.1080/00273171.2014.931800.

\bibitem{brand2016ipace}
M.~Brand, K.~S.~Young, C.~Laier, K.~W\"olfling, and M.~N.~Potenza, ``Integrating psychological and neurobiological considerations regarding the development and maintenance of specific internet-use disorders: An Interaction of Person-Affect-Cognition-Execution (I-PACE) model,'' \emph{Neuroscience \& Biobehavioral Reviews}, vol.~71, pp.~252--266, 2016, doi: 10.1016/j.neubiorev.2016.08.033.

\bibitem{lang2000limited}
A.~Lang, ``The limited capacity model of mediated message processing,'' \emph{Journal of Communication}, vol.~50, no.~1, pp.~46--70, Mar. 2000, doi: 10.1111/j.1460-2466.2000.tb02833.x.

\bibitem{hobfoll1989cor}
S.~E.~Hobfoll, ``Conservation of resources: A new attempt at conceptualizing stress,'' \emph{American Psychologist}, vol.~44, no.~3, pp.~513--524, Mar. 1989, doi: 10.1037/0003-066X.44.3.513.

\bibitem{singh2025attention}
N.~Singh and B.~Prusty, ``The impact of problematic short-form media usage on attention control and delayed gratification in young adults,'' \emph{International Journal of Interdisciplinary Approaches in Psychology}, vol.~3, no.~5, pp.~1248--1256, 2025, doi: 10.37939/jnah.v3i03.172.


\bibitem{review2026shortvideo}
A.~E.~Mona, V.~Roshith, R.~Peter, P.~Roy, A.~Hassan, M.~Devika, and M.~Trishala, ``Short video addiction and its impact on cognitive functioning in adolescents and youth: A systematic review,'' \emph{International Journal of Adolescence and Youth}, vol.~31, no.~1, Art. no.~2623337, 2026, doi: 10.1080/02673843.2026.2623337.

\bibitem{shortvideo2024physical}
Z.~Zhao and Y.~Kou, ``Effects of short video addiction on college students' physical activity: The chain mediating role of self-efficacy and procrastination,'' \emph{Frontiers in Psychology}, vol.~15, art.~1429963, 2024, doi: 10.3389/fpsyg.2024.1429963.

\bibitem{curran1996normality}
P.~J.~Curran, S.~G.~West, and J.~F.~Finch, ``The robustness of test statistics to nonnormality and specification error in confirmatory factor analysis,'' \emph{Psychological Methods}, vol.~1, no.~1, pp.~16--29, 1996.

\bibitem{cronbach1951alpha}
L.~J.~Cronbach, ``Coefficient alpha and the internal structure of tests,'' \emph{Psychometrika}, vol.~16, no.~3, pp.~297--334, 1951, doi: 10.1007/BF02310555.

\bibitem{mcdonald1999test}
R.~P.~McDonald, \emph{Test Theory: A Unified Treatment}. Mahwah, NJ, USA: Lawrence Erlbaum, 1999, doi: 10.4324/9781410601087.

\bibitem{kaiser1974kmo}
H.~F.~Kaiser, ``An index of factorial simplicity,'' \emph{Psychometrika}, vol.~39, no.~1, pp.~31--36, 1974, doi: 10.1007/BF02291575.

\bibitem{horn1965parallel}
J.~L.~Horn, ``A rationale and test for the number of factors in factor analysis,'' \emph{Psychometrika}, vol.~30, no.~2, pp.~179--185, 1965, doi: 10.1007/BF02289447.


\bibitem{fornell1981ave}
C.~Fornell and D.~F.~Larcker, ``Evaluating structural equation models with unobservable variables and measurement error,'' \emph{Journal of Marketing Research}, vol.~18, no.~1, pp.~39--50, 1981, doi: 10.2307/3151312.

\bibitem{meade2012careless}
A.~W.~Meade and S.~B.~Craig, ``Identifying careless responses in survey data,'' \emph{Psychological Methods}, vol.~17, no.~3, pp.~437--455, 2012.

\bibitem{cohen1988power}
J.~Cohen, \emph{Statistical Power Analysis for the Behavioral Sciences}, 2nd~ed. Hillsdale, NJ, USA: Lawrence Erlbaum, 1988.

\end{thebibliography}
\end{document}